\documentclass[10pt,twocolumn,aps,pra,showpacs,superscriptaddress,floatfix,longbibliography]{revtex4-1}

\usepackage[T1]{fontenc}
\usepackage{graphicx}
\usepackage{amssymb}
\usepackage{amsmath}
\usepackage{color}
\usepackage{psfrag}
\usepackage{epsfig}
\usepackage{times}
\usepackage{bbm}
\usepackage{bm}
\usepackage{hyperref}
\usepackage[normalem]{ulem}
\usepackage{amsthm}
\usepackage{comment}
\usepackage{epstopdf}
\usepackage{graphicx}
\usepackage{stackengine,xcolor}
\usepackage{tcolorbox}
\usepackage{makecell}

\newcommand{\je}[1]{{\color{black} #1}}

\definecolor{nred}{rgb}{0.9,0.1,0.1}
\definecolor{nblack}{rgb}{0,0,0}
\definecolor{nblue}{rgb}{0.2,0.2,0.8}
\definecolor{ngreen}{rgb}{0.2,0.6,0.2}

\newcommand{\newtext}[1]{{\color{black} #1}}

\newcommand{\blu}{\color{black}}

\usepackage{etoolbox}
\usepackage{bbold} % use \mathbb{1} to produce \openone

\newcommand{\ket}[1]{| #1 \rangle}
\newcommand{\bra}[1]{\langle #1 |}

\newcommand{\beq}{\begin{gray}}
\newcommand{\eeq}{\end{eqnarray}}

\newcommand{\Aone}{{\rm A}_1}
\newcommand{\Atwo}{{\rm A}_2}
\newcommand{\Bone}{{\rm B}_1}
\newcommand{\Bonebar}{\bar{{\rm B}}_1}
\newcommand{\JJa}{\mathcal{J}_{a}^{\Aone\rightarrow\Atwo}}
\newcommand{\JJax}{\mathcal{J}_{a|x}^{\Aone\rightarrow\Atwo}}
\newcommand{\JJaxs}{\mathcal{J}_{a|x}}
\newcommand{\rhoA}{\rho^{\Aone}}
\newcommand{\Eby}{E_{b|y}^{{\rm B}_1}}
\newcommand{\LAB}{\Lambda^{{\rm A}_2\rightarrow{\rm B}_1}}
\newcommand{\IIax}{\mathcal{I}_{a|x}^{{\rm A}_1\rightarrow{\rm B}_1}}
\newcommand{\IIaxs}{\mathcal{I}_{a|x}}

\newcommand{\EBB}{\mathcal{E}}

\DeclareMathOperator{\tr}{tr}

\theoremstyle{definition}

\AtBeginDocument{%
    \newwrite\bibnotes
    \def\bibnotesext{Notes.bib}
    \immediate\openout\bibnotes=\jobname\bibnotesext
    \immediate\write\bibnotes{@CONTROL{REVTEX41Control}}
    \immediate\write\bibnotes{@CONTROL{%
    apsrev41Control,author="08",editor="1",pages="1",title="0",year="1"}}
     \if@filesw
     \immediate\write\@auxout{\string\citation{apsrev41Control}}%
    \fi
}%

\begin{document}

\title{(Semi-)device independently 
characterizing quantum temporal correlations}

\author{Shin-Liang Chen}
\affiliation{Department of Physics, National Chung Hsing University, Taichung 402, Taiwan}
\affiliation{Physics Division, National Center for Theoretical Sciences, Taipei 106319, Taiwan}
\affiliation{Center for Quantum Frontiers of Research \& Technology (QFort), National Cheng Kung
University, Tainan 701, Taiwan}

\author{Jens Eisert}
\affiliation{Dahlem Center for Complex Quantum Systems, Freie Universit\"at Berlin, 14195 Berlin, Germany}

\date{ \today}

\begin{abstract}
We develop a framework for characterizing quantum temporal correlations in a general temporal scenario, in which an initial quantum state is measured, sent through a quantum channel, and finally measured again. This framework does not make any assumptions on the system nor on the measurements, namely, it is device-independent. It is versatile enough, however, to allow for the addition of further constraints in a semi-device-independent setting. Our framework serves as a natural tool for quantum certification in a temporal scenario when the quantum devices involved are uncharacterized or partially characterized. It can hence also be used for characterizing quantum temporal correlations when one assumes an additional constraint of no-signalling in time, there are upper bounds on the involved systems' dimensions, rank constraints -- for which we prove genuine quantum separations over local hidden variable models -- or further linear constraints. We present a number of applications, including bounding the maximal violation of temporal Bell inequalities, quantifying temporal steerability, bounding the maximum successful probability in quantum randomness access codes.
\end{abstract}

\maketitle
%%%%%%%%%%%%%%%%%%%%%%%%%%%%%%%%%%%%%%%%%%%%%%%%%%%%%%%%%%%%%%%%%%%%%%%%%
%\sBox{\begin{enumerate}
%\item Bounds based on Bell violation
%\end{enumerate}}

Bell's theorem~\cite{Bell64} limits correlations that classical local- hidden-variable models exhibits. This feature of quantum mechanics, referred to as \emph{non-locality}~\cite{Brunner14}, is not only the defining feature that sets apart quantum from classical mechanics, but also exploited in technological-minded applications. Notably, it can be used in new modes of quantum certification not requiring any (possibly unwarranted) assumptions on the states nor on the measurements. In such \emph{device-independent} (DI) quantum certification~\cite{Acin07,Scarani12,Brunner14,BenchmarkingReview}, interestingly, data alone can be seen as being sufficient to certify properties. Along %these 
\je{this}
line of thought,
randomness certification~\cite{Pironio10},  entanglement verification~\cite{Liang15,Baccari17}
and estimation \cite{Moroder13}, quantum state cerification~\cite{Yang14}, 
steerability witnessing \cite{CBLC16,CBLC18}, 
and measurement incompatibility certification \cite{CMBC2021}
%These results 
have all been obtained through the observed non-local 
correlations only and no assumption is made on the shared 
quantum state nor the measurement involved. The \emph{Navascu{\'e}s-Pironio-Ac{\'i}n} 
%(NPA) not used!
hierarchy~\cite{NPA,NPA2008,Doherty08,Moroder13} -- \je{building on earlier
work~\cite{Doherty04,Hierarchies}}
-- has been 
%an instrumental 
\je{a key} tool in these efforts. 
The framework of device independence is compelling, in that one learns about properties of
quantum systems without making assumptions about the devices with which these properties
are being assessed.

That said, the original Bell scenario referring to spatial correlations is by no means the only setting that 
certifies quantum features beyond what local-hidden-variable models can deliver. It has been 
extended to include temporal correlations,
making reference to non-macro-realistic temporal correlations of single systems between two instances in time~\cite{Emary2013,Vitagliano2022}. 
Leggett and Garg~\cite{Leggett85} showed that, in quantum theory, there exist non-macrorealistic temporal correlations. The original Leggett-Garg scenario is as follows: A quantum state is initially prepared and sent through a quantum channel. During the dynamics, the same measurement is performed at some, at least three, points in time.
\je{This}
%Such a scenario 
has then been generalized to an identical 
preparation step, but followed by multiple choices of measurements at each point of time~\cite{Brukner2004,Fritz2010}. Such a setting 
is dubbed \emph{temporal Bell scenario}, since one may view it as a temporal analogue of the standard Bell scenario. Unlike the Leggett-Garg scenario, measurement outcomes between \emph{two} points of time are sufficient to observe non-macrorealistic correlations. Like the situation in the Bell scenario, researchers are searching for a practical way to characterize quantum temporal correlations.  The question is, given observed statistics in a temporal scheme, do there exist %any 
\je{quantum states} and measurements reproducing such statistics? Steps have been taken to characterize quantum temporal correlations in the standard Leggett-Garg scenario \cite{Budroni13}. Nevertheless, characterizing quantum temporal correlations in the temporal Bell scenario remains an open problem, again with implications for device-independence. Indeed, it \je{is not even} known whether such an approach can be pursued at all. From the practical point of view, temporal correlations play an essential role in modern quantum technologies. Famous instances include unitary evolution in quantum circuits and Bennett-Brassard~\cite{Bennett2014} type quantum key distribution. Therefore, studying and characterizing temporal correlations advances the implementation of these cutting-edge technologies. Moreover, since many of them involve the issue of information security, providing a (semi)device-independent framework renders 
them more practical or equips them with more stringent
security promises.

In this work, we develop a framework called \emph{instrument moment matrices} (IMMs) to characterize quantum temporal correlations in a temporal Bell scenario. The IMMs are matrices of expectation values of the post-measurement states, where measurements are described by \emph{instruments}. By construction, if the initial state and the measurements follow quantum theory, the IMMs are positive semi-definite. As such,  quantum temporal correlations can be characterized by semi-definite programming \cite{BoydBook}. Besides, the characterization will be more accurate when the size of IMMs becomes larger (see \cite{NPA,NPA2008} for the original idea behind such a hierarchical characterization and
\cite{Moroder13,Lang2014,Berta16,CBLC16,CBLC18,Bowles2020,CMBC2021,Lin2022} for some 
variants). Our characterization is implemented both 
in a fully \emph{device-independent} (DI) and \emph{semi-DI}
%\footnote{The term ``semi-DI'' normally refers to the scenario where the dimension of the system is assumed in a DI scheme. In our work, instead, we refer to a general situation where there are additional assumptions made on the system in a DI scheme.} 
fashion that incorporates partial knowledge about the devices: \je{We
generalize the reading of semi-DI settings of Ref.~\cite{Liang01}
and} advocate---complementing similarly motivated steps 
\je{closer to the setting of fully specified devices
of ``semi-device-dependent'' 
characterization}
%characterized devices
%along the lines of
\cite{Blind}---that this \emph{intermediate regime} is highly reasonable and important.
%(building up on earlier ideas along these lines \cite{Blind}). 
By DI we mean that the results are based on the \emph{observed}  temporal correlations only, but no measurements and channels have to be specified a-priori. In the temporal scenario, there is no way to rule out the possibility of sending information 
from an earlier time; therefore, we assume there are no side channels in our setting.
%In other words, we assume that we are not in an adversarial scenario such as in that of quantum key distribution.
However, since the space of temporal correlations is so abundant that temporal quantum correlations can, in general, be realized by classical ones~\cite{Clemente2016,Brierley2015},
% (see Ref.~\cite{Brierley2015} from the perspective of communication complexity), 
we have to add additional constraints to reveal quantum advantages. 
%\je{[This is unclear to me.]} \red{[SL: In the temporal scenario considered here, if there is no restriction on the communication from the earlier observer to the later one, then classical strategies can always reproduce the quantum strategy. However, if the communication ability is restricted, there exist some correlations that cannot be reproduced by classical strategies. This is what I would like to convey. See also the paragraph above the blue sentences on the left column of page 3 for another description.]}
For this reason, we further consider 1) the constraint of \emph{no-signaling in time}, 2) the constraint on the system's dimension, and 3) the constraint on the system's rank. We show that IMMs allow us to characterize several quantum resources and tasks in DI and semi-DI scenarios. These include computing the maximal quantum violation of temporal Bell inequalities, estimating the degree of temporal steerability, computing the successful probabilities in scenarios of quantum randomness access codes,  and identifying quantum state preparation.
For including the rank constraint, to the best of our knowledge, this is the first work to enforce additional constraint apart from the dimensional constraint into a device-independent scenario~\cite{FN3}.

\begin{figure}
\includegraphics[width=0.6\linewidth]{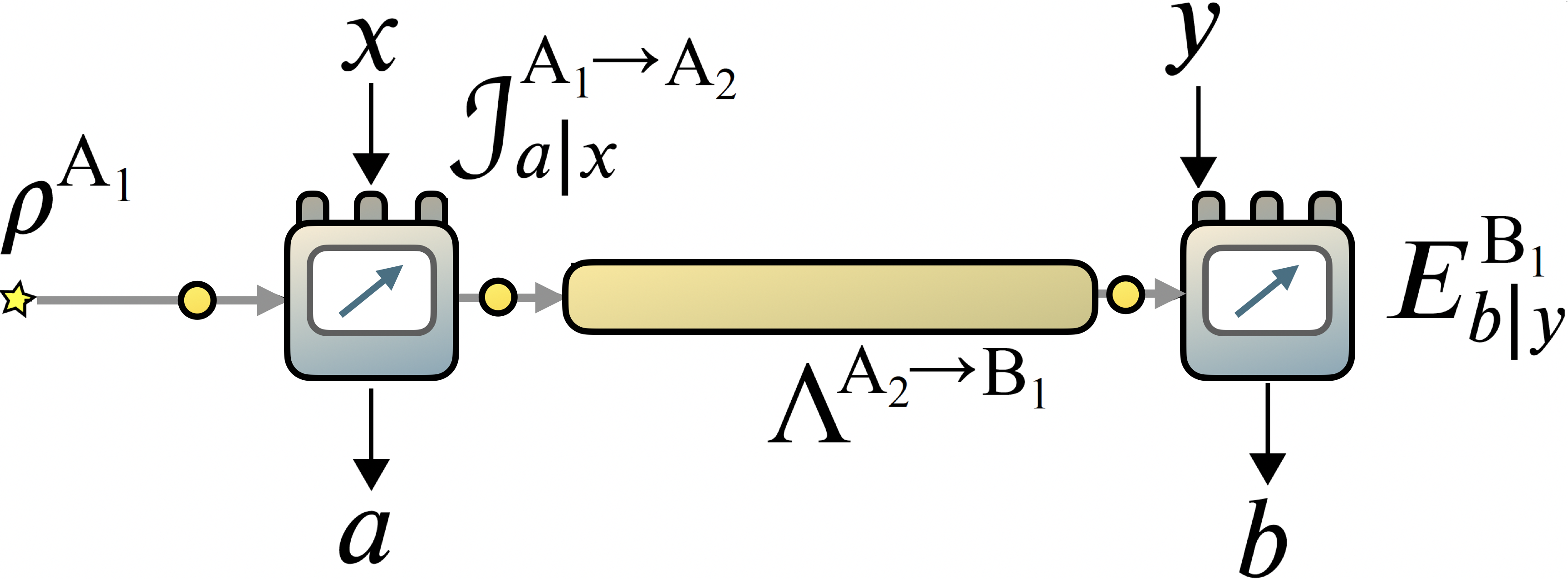}
\caption{The scenario considered in this work.% See the main text for the detailed description.
%
%During each round, the prepared initial quantum state $\rhoA$ undergoes a general quantum measurement described by an element of the set of instruments $\JJax$. The post-measurement state $\JJax(\rhoA)$ is then sent through a quantum channel $\LAB$. The final state $\LAB[\JJax(\rhoA)]$ is measured described by a POVM element $\Eby$. Without knowing further details, the only accessible data is the observed probabilities $P(a,b|x,y)$ as can be estimated by repeating many rounds of the setup.
}
\label{Fig_scenario}
\end{figure}

\textit{The scenario.}
First, we introduce the %formulation of 
notion of an \emph{instrument}. An instrument $\{\JJa\}$
%: $\mathcal{L}(\mathcal{H}_{\Aone})\rightarrow\mathcal{L}(\mathcal{H}_{\Atwo})$
is a set of 
\emph{completely positive} and trace non-increasing maps mapping a quantum state $\rhoA$ to a post-measurement state $\JJa(\rhoA)$ where $a\in\mathcal{A}=\{0,1,2,\dots \}$ can be treated as the assigned outcome associated with the state $\JJa(\rhoA)$. The probability of obtaining the outcome $a$, denoted by $P(a)$, can be computed via $P(a)=\tr(\JJa(\rhoA))$, therefore, one has $\tr\sum_a \JJa(\rhoA)=\tr(\rhoA)$ due to the normalization.

In our scenario, we can choose different instruments to measure the state. %Therefore, 
%Consequently, 
We use the notation $\{\JJax\}$ to denote the collection of instruments, where $x\in\mathcal{X}=\{0,1,2,\dots \}$ %is a real number labeling 
labels the choice of measurement settings (see Fig.~\ref{Fig_scenario}). The post-measurement state $\JJax(\rhoA)$ is then submitted into a quantum channel $\LAB$.
%: $\mathcal{L}(\mathcal{H}_{\Atwo})\rightarrow\mathcal{L}(\mathcal{H}_{\Bone})$. 
Finally, the evolved state is measured by another measurement. At this stage, we only care about the 
%measurement 
outcome, %therefore 
and hence the measurements can be described by \emph{positive operator-valued measures} (POVMs) $\{\Eby\}$
that are positive semi-definite 
%which satisfies the positive semi-definiteness 
$\Eby\succeq 0$ and
normalized \je{as}
%normalization 
$\sum_b\Eby=\openone$, 
where $b\in\mathcal{B}$ and $y\in\mathcal{Y}$ denote the measurement outcome and setting, respectively. By repeating the above experiment, we observe a set of probabilities $\{P(a,b|x,y):=P(b|a,x,y)P(a|x)\}$, 
\je{termed} %a 
\emph{temporal correlations}, which can be obtained by the Born rule
\begin{eqnarray}
P(a,b|x,y) &=& \tr\left\{ \Eby \left[\LAB \left(\JJax(\rhoA)\right) \right]  \right\}
\nonumber
\\
&=&\tr\bigl[\Eby \IIax(\rhoA) \bigr]
\label{Eq_Pabxy_Born}
\end{eqnarray}
\je{where} %in the second equality, 
$\{\IIax:=\LAB\circ\JJax\}_a$ is %still
\je{a} valid instrument for each $x$.
%In a temporal scenario, there exists an inherent constraint that a futural observer can not send any information to the past, i.e., the constraint of \emph{arrow of time}, yielding $\sum_b P(a,b|x,y) = \sum_b P(a,b|x,y')$ for all $y\neq y'.$

%
%
%\begin{figure*}
%\begin{minipage}[c]{.49\textwidth}
%\includegraphics[width=7cm]{Fig_I3622_type.pdf}\\
%\text{(a)}
%%\label{Fig_I3622_type}
%\end{minipage}
%\begin{minipage}[c]{.49\textwidth}
%\includegraphics[width=8.5cm]{Fig_DI_fidelity_EBI_and_I3622.pdf}\\
%\quad\\
%\text{(b)}
%%\label{Fig_fidelity_bound_I3622}
%\end{minipage}
%\caption{
%(a) 
%}
%\label{Fig_EB_and_I3622}
%\end{figure*}

\textit{The instrument moment matrices and their DI formulation.}
The \emph{instrument moment matrices} (IMMs) are constructed by applying complete-positive maps $\EBB$
%: $\mathcal{L}(\mathcal{H}_{\Bone})\rightarrow\mathcal{L}(\mathcal{H}_{\Bonebar})$ 
on the post-measurement states $\IIax(\rhoA)$, i.e., $\EBB(\IIax(\rhoA))=\sum_n K_n[\IIax(\rhoA)] K_n^\dag$ with $K_n:=\sum_i\ket{i}_{\Bonebar \Bone}\bra{n}S_i$ being the Kraus operators. Here, $\{\ket{i}_{\Bonebar}\}$ and $\{\ket{j}_{\Bone}\}$ are orthonormal bases for the output space and input space, respectively. Following Ref.~\cite{Moroder13}, given a level $\ell$ we choose $\{S_i\}$ as $\openone\cup\mathcal{S}^{(1)}\cup\mathcal{S}^{(2)}\cup \dots \cup\mathcal{S}^{(\ell)} $, where $\mathcal{S}^{(\ell)}$ is composed of the $\ell$th order products of the operators in the set $\{\Eby\}_{b=1,\dots, |\mathcal{B}|-1}^{y=1,\dots ,|\mathcal{Y}|}$. The $\ell$th-level IMMs can be defined as
\begin{equation}
\chi_{a|x}^{(\ell)} := \EBB[\IIaxs(\rhoA)] = \sum_{i,j}\ket{i}\bra{j}\tr\left[ \IIaxs(\rhoA) S_j^\dag S_i \right].
\label{Eq_IMM}
\end{equation}
Therefore, the entry of the $i$th row and $j$th column of $\chi_{a|x}^{(\ell)}$ can be treated as the ``expectation value'' of the product of $S_j^\dag$ and $S_i$ given the state $\IIax(\rhoA)$.
In Appendix \ref{SecApp_example_IMM}, we explicitly provide 
an example of IMMs.
Note that the IMMs are positive semi-definite whenever $\IIaxs$, $\rho$, $\Eby$ are quantum realizable: 
%In other words, 
The constraints of positive semi-definiteness $\chi_{a|x}^{(\ell)}\succeq 0$ serve as a natural characterization of the quantum set of temporal correlations $\{P(a,b|x,y)\}$. The characterization is improved when the level $\ell$ increases. When the improvement is hard to be observed from a level $\ell_c$, we say $\chi_{a|x}^{(\ell_c)}$ provides a proper approximation of the quantum set of temporal correlations. %In the rest of paper 
We will from now on use the notation $\chi_{a|x}$ to simply denote $\chi_{a|x}^{(\ell)}$. 
%when there is no risk of confusion. 

%In general, one can decompose IMMs into the characterized and unknown parts:
%\begin{equation}
%\chi_{a|x} = \chi_{a|x} 
%\end{equation}

When focusing on temporal correlations, quantum systems do not ``outperform'' classical systems
%. The reason is 
in that a classical system with a sufficiently high dimension carries information allowing observers at later time to obtain. The %most 
simplest scheme is that an observer at earlier time can just send all the information about the measurement settings and outcomes to an observer at later time, then the correlation space will be filled by such a strategy. To let quantum systems demonstrate their superior performance, a constraint is to limit their dimension. By doing so, it has been shown that quantum systems outperform classical systems with the same dimension~\cite{Gallego10}. If we require that the entire system is embedded in dimension \emph{at most} $d$,  we have
$P(a,b|x,y) = \tr
\{ \Eby [ \IIax(\rhoA) ]  \},$
with $\rhoA\in\mathcal{L}(\mathcal{H}_d^{{\rm A}_1})$, $\IIax: \mathcal{L}(\mathcal{H}_d^{{\rm A}_1})\rightarrow \mathcal{L}(\mathcal{H}_d^{{\rm B}_1})$, and $\Eby\in \mathcal{D}(\mathcal{H}_d^{{\rm B}_1})$. Following the idea of Ref.~\cite{Navascues15prl}, the set of probabilities $P(a,b|x,y)$ generated by $d$-dimensional systems can be characterized by embedding IMMs into dimension-restricted IMMs, namely,
%\begin{equation}
$\{\chi_{a|x}\}_{a,x}\in\mathcal{G}_d$
%\end{equation}
where $\mathcal{G}_d$ is the set of IMMs composed of $d$-dimensional quantum systems.

The second kind of constraints we would like to impose is an upper bound on the rank of Bob's measurements. To this end, when generating Bob's $d$-dimensional POVMs $\Eby$, we generate $\Eby$ with rank $k$ only, namely, Rk$(\Eby)=k,$ where Rk$(\cdot)$ denotes the rank.
%is the rank of the operator. 
We denote \je{with} $\mathcal{G}_d^k$ %as 
the set of IMMs with such a construction, i.e., $\{\chi_{a|x}\}_{a,x}\in\mathcal{G}_d^k$. 
%Note that 
In our method, the rank constraint cannot be considered alone without the dimensional constraint. The reason is that when generating the POVM elements $\Eby$, the dimension of them is automatically defined. In the same sense, in the typical dimension-constraint scenario, one implicitly 
sets the upper bound on the rank of measurements to be full rank.
The final constraint we would like to consider is the so-called \emph{no signaling in time} (NSIT). Such a constraint states that the observer at earlier time cannot transmit information by changing the measurement settings, i.e., $\sum_a P(a,b|x,y)=\sum_a P(a,b|x',y)$, yielding $\sum_a \chi_{a|x} = \sum_a \chi_{a|x'}~\forall x\neq x'.$
Since no information is transmitted between two observers at different points of time, the NSIT constraint in the temporal scenario is in general the same as the typical (i.e., spatial) Bell scenario.

Now we have four types of constraints used for characterizing quantum sets of temporal correlations: the device-independent (DI), DI $+$ dimensional, DI $+$ rank, and NSIT constraints. They are respectively %summarized and 
denoted as
%follows:
%\begin{tcolorbox}
\begin{itemize}
\item DI: $\chi_{a|x}\succeq 0$ %$\forall a,x$.
\item {DI$+$Dim.}: $\chi_{a|x}\succeq 0$, $\{\chi_{a|x}\}_{a,x}\in\mathcal{G}_d$.
\item {DI$+$Dim.$+$Rank:} $\chi_{a|x}\succeq 0$,
$\{\chi_{a|x}\}_{a,x}\in\mathcal{G}_d^k$.
\item NSIT: $\chi_{a|x}\succeq 0$, $\sum_a \chi_{a|x}=\sum_a \chi_{a|x'}$ $\forall x\neq x'$.
\end{itemize}
%\end{tcolorbox}
%For convenience, 
% When we mention \emph{semi-device-independent} (semi-DI) scenarios, we include the second to fourth types of constraints.

\textit{{Quantum upper bounds on temporal Bell inequalities.}}
To demonstrate that the IMMs provide a proper characterization, we first show that the IMMs can be used to compute an upper bound on the maximal quantum violation of a temporal Bell inequality. \newtext{This result is also crucial from the practical point of view since we have to make sure that the temporal Bell inequality used for certifying non-classicality (i.e., a nonmacrorealistic dynamics~\cite{Emary2013}) provides different bounds for quantum and classical models.} To simplify the problem, we consider the temporal 
\emph{Clauser-Horne-Shimony-Holt} (CHSH) scenario~\cite{Clauser69,Brukner2004,Zela2007,Fritz2010}, 
i.e., the scenario with binary settings and outcomes. The generalization to arbitrary scenarios can be straightforwardly obtained. The temporal CHSH inequality is written as
\begin{equation}
K_{\rm CHSH}:=\langle A_0 B_0\rangle + \langle A_0 B_1\rangle + \langle A_1 B_0\rangle - \langle A_1 B_1\rangle\leq 2,
\end{equation}
where $\langle A_x B_y\rangle:=P(a=b|x,y)-P(a\neq b|x,y)$. The bound with the value of $2$ is obtained from the so-called \emph{macrorealistic model}~\cite{Emary2013,Vitagliano2022}. As been known, the inequality can be violated since quantum physics does not admit a macrorealistic model. An quantum upper bound on the inequality can be computed via \emph{semi-definite programming}~\cite{BoydBook}
\begin{equation}
\max\{ K_{\rm CHSH}\vert \chi_{a|x}\succeq 0,~\forall a,x\}.
\label{Eq_QB_KCHSH}
\end{equation}
The solution gives us the value of $4$, \je{the maximal algebraic value}. This coincides with one of results in \cite{Hoffmann2018}, which states that any correlation admitting the arrow of time can always be realized by quantum theory~\cite{FN1}. Even when we consider the dimensional constraint, i.e., the DI+Dim. constraint with $d=2$, the tight quantum upper bound on $K_{\rm CHSH}$ is still $4$. The bound is tight since there exists a quantum realization to achieve the bound. It is interesting to note that if we further restrict Bob's POVMs to be rank $1$, i.e., the DI+Dim.+Rank constraint with $(d,k)=(2,1)$, the upper bound on $K_{\rm CHSH}$ will be within the numerical precision with $2\sqrt{2}$, same with the Tsirelson bound~\cite{Tsirelson1980} in the spatial CHSH scenario. Finally, if we consider the NSIT constraint, the scenario will be the same as that of the spatial CHSH; that is, two-way communication is forbidden. The upper bound on $K_{\rm CHSH}$ we obtain is within the numerical precision with the Tsirelson bound~\cite{Tsirelson1980}, $2\sqrt{2}$. \newtext{The different quantum bounds for the latter two with the former two schemes provide an important application: Exceeding the value of $2\sqrt{2}$ \emph{sufficiently identifies} at least one of the following three facts: 1) the underlying qubit measurements are not one rank (i.e., full rank), 2) the dimension of the system is beyond qubit, and 3) there exists one-way communication.}

\textit{{Bounding the degree of temporal steerability.}}
The idea of temporal steerability was first proposed in Ref.~\cite{YNChen14}. 
%The authors showed that, under the assumption of non-invasive measurement of the earlier point of time, there exists a temporal analogue of a steering inequality~\cite{Cavalcanti09}, while quantum theory can violate such a temporal steering inequality. 
The works of Refs.~\cite{SLChen16_2,SLChen17,Li15} have reformulated the classical model in \cite{YNChen14} by introducing the hidden-state model~\cite{Wiseman07}.
In our formulation, the hidden-state model is described by (see also Ref.~\cite{Uola2018}):
$\IIaxs(\rho)=\sum_\lambda P(\lambda)P(a|x,\lambda)\sigma_\lambda$, where $P(\lambda)$, $P(a|x,\lambda)$ are probabilities and $\sigma_\lambda$ are quantum states. The equation above tells us that the post-measurement states $\IIaxs(\rho)$ are simply a classical post-processing of the set of fixed states $\sigma_\lambda$. In quantum theory, there exist instruments $\IIaxs$ such that the post-measurement states $\IIaxs(\rho)$ do not admit a hidden-state model. The incompatibility with a hidden-state model is called \emph{temporal steering}.
%, and the degree of which can be measured by the \emph{temporal steering robustness}~\cite{HYKu16}.
%
Here, we show that by observing the statistics $P(a,b|x,y)$, we are still capable of bounding the degree of temporal steerability in DI and semi-DI scenarios. See Appendix~\ref{SecApp_DITSR} for the detailed derivation and computational results. \newtext{Very recently, it has been shown that temporal steerability has a physical meaning: it is equivalent to the time a thermodynamic bath requires to bring the states $\IIaxs(\rho)$ to the thermal state~\cite{CYHsieh2024}. Therefore, our method can also be used for device-independently estimating the thermalization time.}

\begin{figure}
\includegraphics[width=0.52\linewidth]{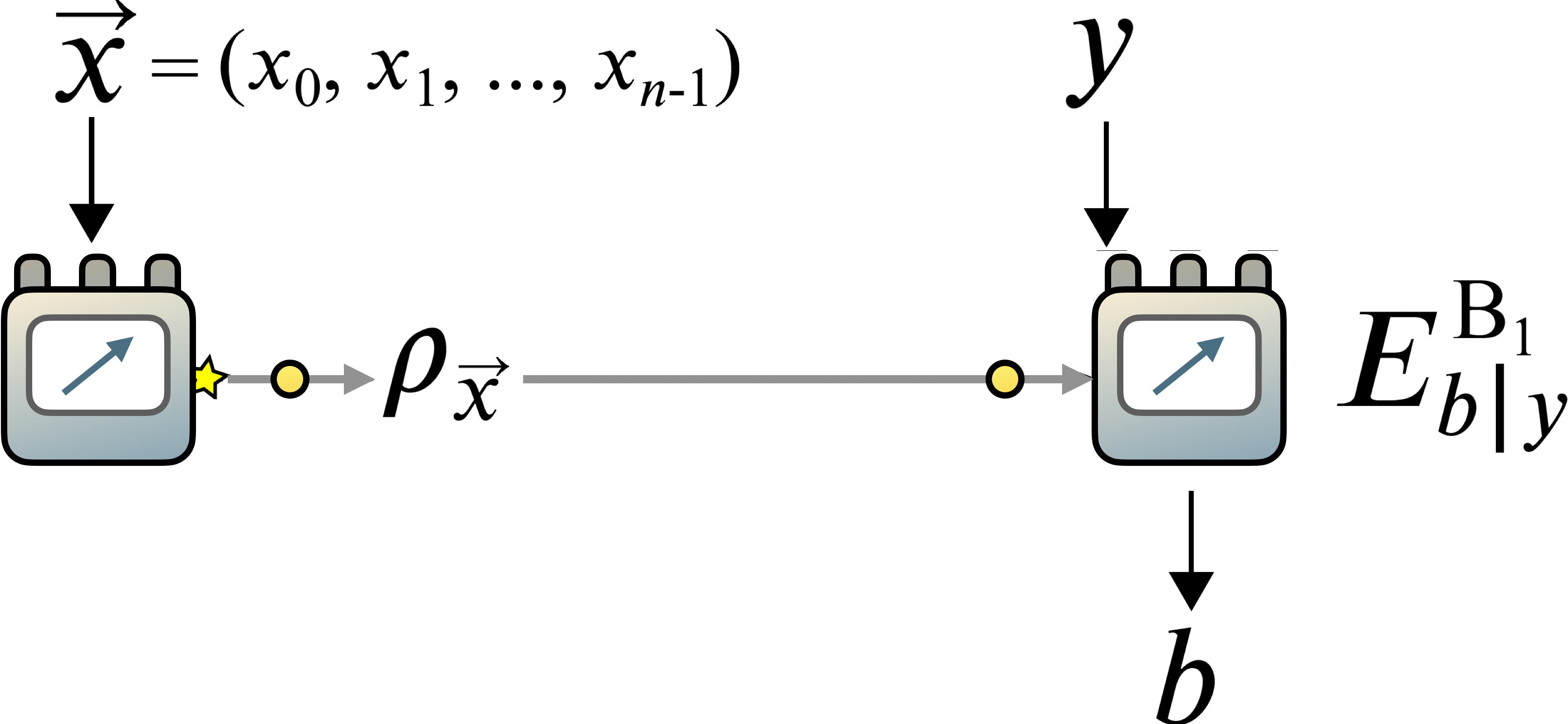}
\caption{The $n\rightarrow 1$ \emph{quantum randomness access codes} (QRACs).
}
\label{Fig_QRAC}
\end{figure}

\textit{{Characterization 
%of successful probabilities 
\je{of} quantum randomness access codes.}}
In the $n\rightarrow 1$ \emph{random access code} (RAC) scenario, an observer (Alice) has $n$ bits of information, denoted by $\vec{x}=(x_0,x_1,\dots x_y,\dots ,x_{n-1})$ with $x_i\in\{0,1\}$. She then encodes them into a single bit and sends it to the other observer (Bob) who is queried for guessing Alice's $y$th bit. Their goal is to 
%make 
\je{maximize} Bob's guessing probability, i.e., $P(b=x_y|\vec{x},y)$, %as high as possible, 
where $b$ is Bob's guess (see
%. The scenario is depicted in 
Fig.~\ref{Fig_QRAC}). We denote \je{with} $\mathcal{P}_{n\rightarrow 1}^{\rm C}$ the maximum average (over all $x_y$ and $y$) successful probability by a classical strategy. It has been shown that $\mathcal{P}_{2\rightarrow 1}^{\rm C}=\mathcal{P}_{3\rightarrow 1}^{\rm C}=3/4$.
In quantum theory, Alice's $n$ bits of information are encoded in the way of quantum state preparation, i.e., for each given $\vec{x}$, she sends the associated quantum state $\rho_{\vec{x}}$ to Bob. Bob then performs his $y$th quantum measurement, described by a POVM $\{E_{b|y}\}_b$, on the state. The quantum realization of the guessing probability will be $P(b=x_y|\vec{x},y) = \tr(E_{b|y}\rho_{\vec{x}})$. Denoting $\mathcal{P}_{n\rightarrow 1}^{\rm Q}$ as the maximum average successful probability by a quantum strategy, it has been shown that $\mathcal{P}_{2\rightarrow 1}^{\rm Q}=\frac{1}{2}(1+1/\sqrt{2})\approx 0.8536$ and $\mathcal{P}_{3\rightarrow 1}^{\rm Q}=\frac{1}{2}(1+1/\sqrt{3})\approx 0.7887$. %In the following, 
%we show 
In Appendix \ref{SecApp_ProbQRAC}, we show how to use IMMs to recover these quantum bounds.

\textit{{Self-testing quantum states in a prepare-and-measure scenario.}}
Finally, we show that the IMMs can be used for verifying set of quantum states in a semi-DI way. More explicitly, we consider the QRAC scenario in the last section and uniquely (up to some isometries) identify the underlying set of states $\rho_{\vec{x}}$ by the observed probabilities $P(b|\vec{x},y)$ only. Such identification, called \emph{self-testing in a prepare-and-measure scenario}, has been proposed in Refs.~\cite{Tavakoli18b,SHWei2019,Miklin2021universalscheme}. 
%In the following, we 
We here provide 
%another 
\je{an alternative}
approach to achieve the task.
A robust self-testing can be defined as follows 
%(see, e.g.,
%Refs.~
\cite{Tavakoli18b,Supic2020selftestingof}). 
Given an upper bound $d$ on the dimension of the systems involved, we say that the observed correlation $\vec{P}:=\{P(b|\vec{x},y)\}_{b,\vec{x},y}$ robustly self-tests, in a prepare-and-measure scenario, the reference set of states $\vec{\rho}_{\rm ref}:=\{\rho_{\vec{x}}^{\rm ref}\}_{\vec{x}}$ at least with a fidelity $f$ if for each set of states $\vec{\rho}:=\{\rho_{\vec{x}}\in\mathcal{H}_d\}_{\vec{x}}$ compatible with $\vec{P}$ there exists a \emph{completely positive and trace-preserving} map $\Lambda$, such that $F(\vec{\rho}_{\rm ref}, \Lambda(\vec{\rho}))\geq f$. Here, $\Lambda(\vec{\rho})$ represents for $\Lambda(\rho_{\vec{x}})$ for all $\vec{x}$ and $F(\vec{\rho}, \vec{\sigma})$ is the fidelity between two sets of states $\vec{\rho}$ and $\vec{\sigma}$.
%namely~\cite{Liang19}, $F(\vec{\rho}, \vec{\sigma}):=\sum_{\vec{x}} F^{\rm UJ}(\rho_{\vec{x}}, \sigma_{\vec{x}})=\frac{1}{2^n}\sum_{\vec{x}}\tr(\rho_{\vec{x}}\sigma_{\vec{x}}),$ where $F^{\rm UJ}$ is the \emph{Uhlmann-Josza fidelity}~\cite{Uhlmann76,Jozsa1994} and the second equality holds when $\rho_{\vec{x}}$ or $\sigma_{\vec{x}}$ are pure.

\begin{figure}
\includegraphics[width=.75\linewidth]{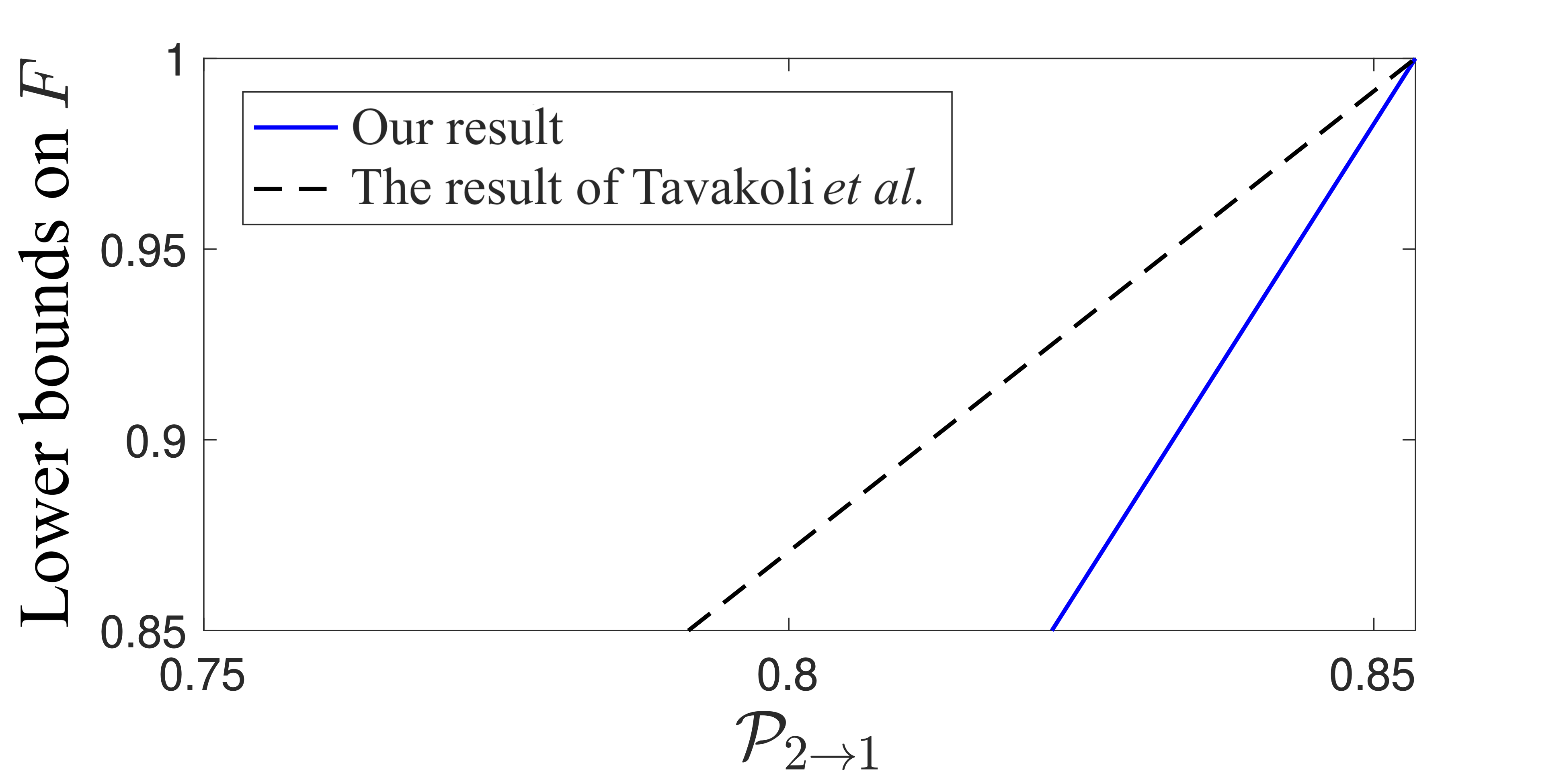}
\caption{Robust self-testing the reference set of states in the prepare-and-measure scenario.}
\label{Fig_min_f_QRAC2to1}
\end{figure}

{\blu
To compute $F(\vec{\rho}_{\rm ref}, \Lambda(\vec{\rho}))$ in a DI way, we use 
%the 
\je{a} method similar \je{to that of} 
%with
Ref.~\cite{Chen2021robustselftestingof}.
%, where the authors self-test \je{steering assemblages}. 
%\je{The idea is to express the \emph{Choi-Jamio{\l}kowski} (CJ) matrix reflecting the channel} in terms of Bob's observables. 
\je{The fidelity can then}
be written as a polynomial where each monomial is of the form $\tr(\rho_{\vec{x}}S_j^\dag S_i)$ with $S_i$ being Bob's observables or their products (see Appendix \ref{SecApp_DI_fidelity}). % Finally, 
 \je{Given} the observed correlation $\vec{P}$, a DI bound on $F(\vec{\rho}_{\rm ref}, \Lambda(\vec{\rho}))$, 
%which we denote 
\je{denoted}
as $F^{\rm DI}$, can be computed 
%by introducing constraints on the IMM 
\je{as}
\begin{equation}
\min \bigl\{ F^{\rm DI}(\vec{\rho}_{\rm ref}, \Lambda(\vec{\rho}))
\Big\vert  \chi_{\vec{x}}\succeq 0,\quad \chi_{\vec{x}}\in\mathcal{G}_{d}^{k} \bigr\}.
\label{Eq_F_DI}
\end{equation}
We consider %an 
\je{the} example of \je{a} $2\to 1$ scenario, where the reference preparation is chosen as a unitary equivalent to $\{\ket{0},\ket{1},\ket{+},\ket{-}\}$, implying $d=2$. We 
%also 
\je{assume} the measurement to be projective (as most works \je{do}), 
%on QRAC scenario do), %hence we have 
\je{so that} $k=1$. %Computing Eq.~\ref{Eq_F_DI}, 
The result is presented by the blue-solid line in Fig.~\ref{Fig_min_f_QRAC2to1}. %In the plot, 
\je{The} observed correlation $\vec{P}$ is represented by the average successful probability $\mathcal{P}_{2\rightarrow 1}:=
\frac{1}{8} \sum_{x_0,x_1,y}P(b=x_y|x_0,x_1,y)$. %As can be seen, 
\je{Given the maximal} quantum value of $\mathcal{P}_{2\rightarrow 1}=\mathcal{P}_{2\rightarrow 1}^Q$, we perfectly self-test the reference set of states with fidelity equal to $1$. When $\mathcal{P}_{2\rightarrow 1}$ is below around $0.8232$, we no longer have self-testing statement, since the fidelity is below the classical %(trivial) 
\je{fidelity} $0.8536$ (see Appendix \ref{SecApp_classical_fidelity})
% for the derivation of the classical fidelity).
We also compare our result with the optimal bounds proposed by Tavakoli \emph{et al.}~\cite{Tavakoli18b}.
%The optimal bounds on the fidelity is proposed in Ref.~\cite{Tavakoli18b}, i.e., the black-dashed line in Fig.~\ref{Fig_min_f_QRAC2to1}. It is an open question \je{how} to find the best expression of the CJ matrix to make our bounds optimal.
}

\newtext{\textit{Generalization to multiple-time and many-body scenarios.}}
\newtext{
Our method can be straightforwardly generalized to two scenarios. The first is considering multiple measurements acting on a single system at different time.
%The quantum realization is applying multiple instruments on the initial quantum state before the final quantum measurement is performed. The associated IMMs can thus be constructed and can also be used for characterizing quantum temporal correlations (see Appendix H~\cite{Suppl} for details.)
For the second scenario, if we are interested in the time evolution of a many-body system, we can go back to our original setting depicted in Fig. \ref{Fig_scenario} and replace the initial state with a many-body state.
%Then, to obtain the temporal correlation, the entire system is measured, sent into a channel, and measured again.
For both generalizations, the mathematical constructions of the IMMs are similar to the standard one (see Appendix \ref{SecApp_generalization} for details). The difference is that the Hilbert spaces involved are much larger, yielding large-size IMMs. To reduce the consumption of computational resources, one may introduce techniques used in some previous works~\cite{FN4}.
%(e.g., the symmetric property~\cite{Tura2014}, commutative property~\cite{Baccari17}, positive-semi-definite constraints~\cite{Frerot2022}, etc.)
The comprehensive investigation is beyond the scope of this work, and we leave the deeper exploration of these issues as future research.
}

\textit{{Summary and discussion.}}
In this work, we have %consider 
established a general temporal scenario and develop a method, dubbed as \emph{instrument moment matrices} (IMMs), to characterize quantum temporal correlations generated by such a scenario. The method of IMMs can be implemented in a fully DI scenario, but we can also include additional constraints (such as the dimension and rank of the system) when these information is accessible. Along the side, we 
contribute to advocating to explore the ``room in the middle'' between the (precise, but very
restrictive) DI and device-specific scenarios: In contrast to Ref.~\cite{Blind} which is close
to device-dependence and is hence dubbed \emph{semi-device-dependent}, we are here close to the DI
regime, in the \emph{semi-device-independent} setting. We explicitly provide several DI and semi-DI examples.
%including bounding the maximal value of temporal Bell inequalities and the minimum degree of temporal steerability. Moreover, its variant allows us to compute the maximal successful probability and certify the set of quantum states in a QRAC scenario.

Regarding implementing our protocol experimentally, there may be some loopholes we have to take care of. For instance, if the detection efficiency of the detectors is too low, there exists a classical model which can be used for reproducing the observed data~\cite{Brunner14}. Besides, if the temporal scenario is set for testing local realism (i.e., the Leggett-Garg test), one may meet the clumsiness loophole~\cite{Emary2013} issue. In that case, one can use another type of temporal Bell inequalities, such as those proposed in Ref.~\cite{Wilde2011}.

%We leave some open questions for the future research. 
%Our work invites a number of questions for future research:  Second, since the construction of the IMMs includes the measurements and channels, we expect that the IMMs can be used for certifying properties of quantum measurements and channels, e.g., incompatible measurements or non entanglement-breaking channels. Finally, it is interesting to see if the IMMs can also be used for self-testing a set of complex-valued states. 

\begin{acknowledgments}
%{\it Acknowledgements.}
We thank Nikolai Miklin, Costantino Budroni, Yeong-Cherng Liang, Armin Tavakoli, and Kai-Siang Chen for fruitful discussions. S.-L.~C.~acknowledges the support of the National Science and Technology Council (NSTC) Taiwan (Grant No.~NSTC 111-2112-M-005-007-MY4) and National Center for Theoretical Sciences Taiwan (Grant No.~NSTC 112-2124-M-002-003). J.~E.~acknowledges support by the 
European Research Council (ERC AdG DebuQC), the BMBF (QR.X), \newtext{the DFG (CRC 183, FOR 2724)}, the 
Munich Quantum Valley (K-8), and the Einstein Foundation.
\end{acknowledgments}

%\bibliography{bib_SDI_temporal_correlations}

\begin{thebibliography}{64}
\bibitem [{FN3()}]{FN3}%
  \BibitemOpen
  \href@noop {} {}\bibinfo {note} {We would like to stress that in
  Ref.~\cite{Navascues15prl}, the general idea of characterizing temporal
  correlations has been proposed. The difference is that
  Ref.~\cite{Navascues15prl} has focused on the prepare-and-measure scenario
  while we consider a two-time-measurement scenario (see
  Fig.~\ref{Fig_scenario}). Building on this, we demonstrate several explicit
  applications.}\BibitemShut {Stop}%
\bibitem [{FN1()}]{FN1}%
  \BibitemOpen
  \href@noop {} {}\bibinfo {note} {Note that in their work, the authors
  consider the scenario where the same instruments are performed at each point
  of time, i.e., $\IIax(\openone)=E_{a|x}^{{\rm B}_1}$, which is a special case
  of our scenario.}\BibitemShut {Stop}%
\bibitem [{FN2()}]{FN2}%
  \BibitemOpen
  \href@noop {} {}\bibinfo {note} {We omit the superscripts representing the
  Hilbert spaces the operators acting on when there is no risk of
  confusion.}\BibitemShut {Stop}%
\end{thebibliography}

%merlin.mbs apsrev4-1.bst 2010-07-25 4.21a (PWD, AO, DPC) hacked
%Control: key (0)
%Control: author (8) initials jnrlst
%Control: editor formatted (1) identically to author
%Control: production of article title (0) allowed
%Control: page (1) range
%Control: year (1) truncated
%Control: production of eprint (0) enabled
%

\clearpage
\onecolumngrid
\appendix

\section{Comparison of different approaches of characterizing quantum temporal correlations}

\vspace{\columnsep}
\begin{center}
\begin{table}[h!]
\centering
\caption{Different approaches of characterizing quantum temporal correlations. Some works consider different types of temporal scenarios. The summaries as well as 
features and applications are briefly listed for comparison.}
\begin{tabular}{|l| l| l|}
\hline
Methods & Summary & Features and applications \\ \hline \hline
This work. & \makecell[l]{The method characterizes correlations between\\ two moments of time in a general temporal scenario.} & \makecell[l]{Bounding the temporal Bell inequalities,\\ quantification of temporal steerability,\\ self-testing QRAC states.}  \\ \hline

Budroni \emph{et al.}~\cite{Budroni13}. & \makecell[l]{Characterization of temporal correlations of \\ \emph{sequential measurements} for both the Leggett-Garg \\ type and the Kochen-Specker type.} & \makecell[l]{Bounding Leggett-Garg inequalities, \\ bounding Kochen-Specker inequalities.}  \\ \hline

Hoffmann \emph{et al.}~\cite{Hoffmann2018}. & \makecell[l]{Characterization of two-dimensional quantum \\ temporal correlations for Kochen-Specker \\ sequential measurements.} & \makecell[l]{Qutrit certification in sequential-measurement scheme.} \\  \hline

Gallego \emph{et al.}~\cite{Gallego10}. & \makecell[l]{Characterization of dimension bounded \\ (both classical and quantum) correlations \\ of \emph{prepare-and-measure} scenario.} & \makecell[l]{Dimension (classical and quantum) certification \\ in prepare-and-measure scheme.}  \\ \hline

\end{tabular}
\end{table}\label{TB_MM_Refs}
\end{center}
\vspace{\columnsep}

\section{Instrument in a nutshell}\label{SecApp_instrument}
An instrument can be seen as a generalization of a \emph{positive operator-valued measure} (POVM) for describing a quantum measurement. If we care about the statistics of measurement outcomes only, a POVM is proper for representing the measurement. However, if we also care about the quantum state after the measurement (called the post-measurement state), an \emph{instrument} is a more appropriate way to describe the measurement. An instrument includes information about the measurement outcome and the change of the input state. As a result, it is equivalent to describing an instrument $\{\mathcal{J}_a\}$ with a set of \emph{subchannels} $\{\Lambda_a\}$, where $a$ labels the measurement outcome and $\Lambda_a$ describes the change of the input state when observing the outcome $a$. That is, the post-measurement (unnormalized) state $\rho'_a$ is obtained via
\begin{equation}
\rho'_a = \mathcal{J}_a(\rho) = \Lambda_a(\rho).
\end{equation}
To obtain the normalized post-measurement state, one has to normalize $\rho'_a$, i.e.,
\begin{equation}
\frac{\rho'_a}{\tr(\rho'_a)} = \frac{\mathcal{J}_a(\rho)}{\tr\big[\mathcal{J}_a(\rho)\big]} = \frac{\Lambda_a(\rho)}{\tr\big[ \Lambda_a(\rho) \big]},
\end{equation}
where the term $\tr\big[\mathcal{J}_a(\rho)\big]=\tr\big[ \Lambda_a(\rho) \big]$ is interpreted as the probability of obtaining the measurement outcome $a$.

\section{An example of the IMMs}\label{SecApp_example_IMM}
In this section, we explicitly present an example of IMMs for dichotomic measurement settings and outcomes. For the $1$st level of semi-definite hierarchy, IMMs are $3\times3$ matrices~\cite{FN2}
%\footnote{We omit the superscripts representing the Hilbert spaces the operators acting on when there is no risk of confusion.}
\begin{equation}
\begin{aligned}
\chi_{a|x}=
\begin{pmatrix}
\tr[\IIaxs(\rho)]              & \tr[\IIaxs(\rho)E_{0|0}^\dag]               & \tr[\IIaxs(\rho)E_{0|1}^\dag] \\
\tr[\IIaxs(\rho)E_{0|0}] & \tr[\IIaxs(\rho)E_{0|0}^\dag E_{0|0}] & \tr[\IIaxs(\rho)E_{0|1}^\dag E_{0|0}] \\
\tr[\IIaxs(\rho)E_{0|1}] & \tr[\IIaxs(\rho)E_{0|0}^\dag E_{0|1}] & \tr[\IIaxs(\rho)E_{0|1}^\dag E_{0|1}]
\end{pmatrix}.
\end{aligned}
\end{equation}
In a DI setting, $\IIax$, $\rhoA$, and $\Eby$ are unknown. However, we are still able to access some information about $\chi_{a|x}$. For instance, entries corresponding to $\tr[\IIaxs(\rho)E_{b|y}]$ 
are $P(a,b|x,y)$, which are accessible in a DI scheme. Besides, since every POVM can be obtained from a projective measurement with a higher dimension~\cite{Neumark}, we can treat $\{\Eby\}$ as projective measurements, i.e., $\Eby E_{b'|y}^{\Bone}=\delta_{b,b'}\Eby$. Thus $\chi_{a|x}$, can be written as
\begin{equation}
\begin{aligned}
\chi_{a|x}=
\begin{pmatrix}
P(a|x)         & P(a,0|x,0)               & P(a,0|x,1) \\
P(a,0|x,0)  & P(a,0|x,0)               & \tr[\IIaxs(\rho)E_{0|1}^\dag E_{0|0}] \\
P(a,0|x,1)  & \tr[\IIaxs(\rho)E_{0|0}^\dag E_{0|1}] & P(a,0|x,1)
\end{pmatrix}.
\end{aligned}
\end{equation}
Entries such as $\tr[\IIaxs(\rho)E_{0|1}^\dag E_{0|0}]$ are not accessible in a DI scenario, therefore, they are some unknown complex numbers, denoted by $u_v$.

\section{Results on the quantification of temporal steerability in DI, DI+Dim., DI+Dim.+Rank, and NSIT scenarios}\label{SecApp_DITSR}
\begin{figure*}
\begin{minipage}[c]{.49\textwidth}
\includegraphics[width=9cm]{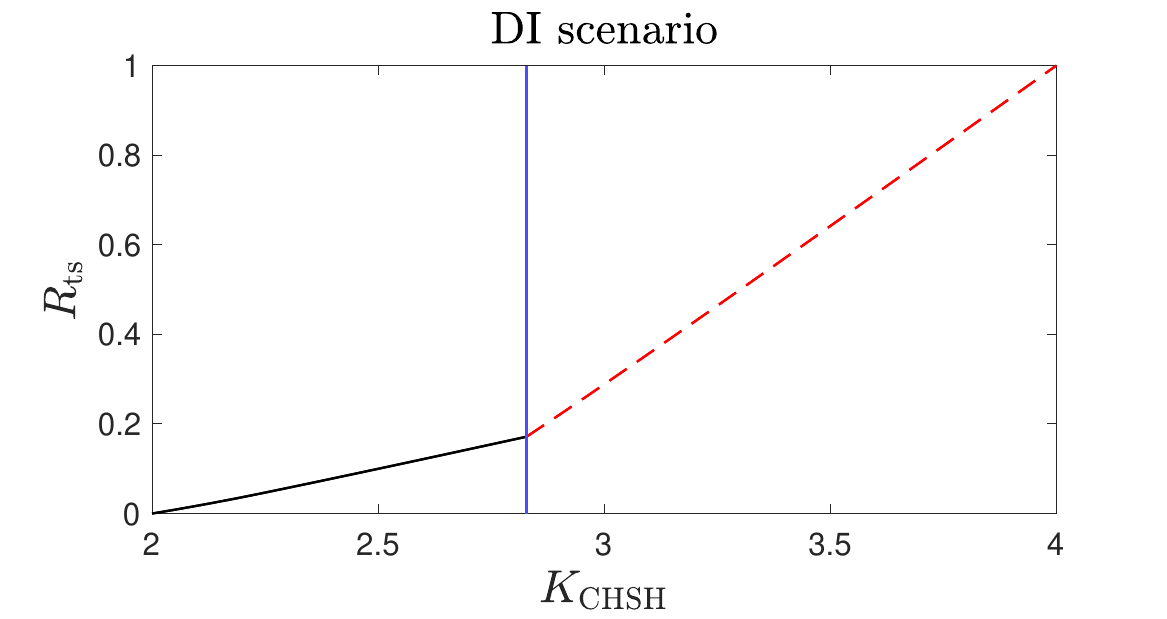}\\
\text{(a)}
\end{minipage}
\begin{minipage}[c]{.49\textwidth}
\includegraphics[width=9cm]{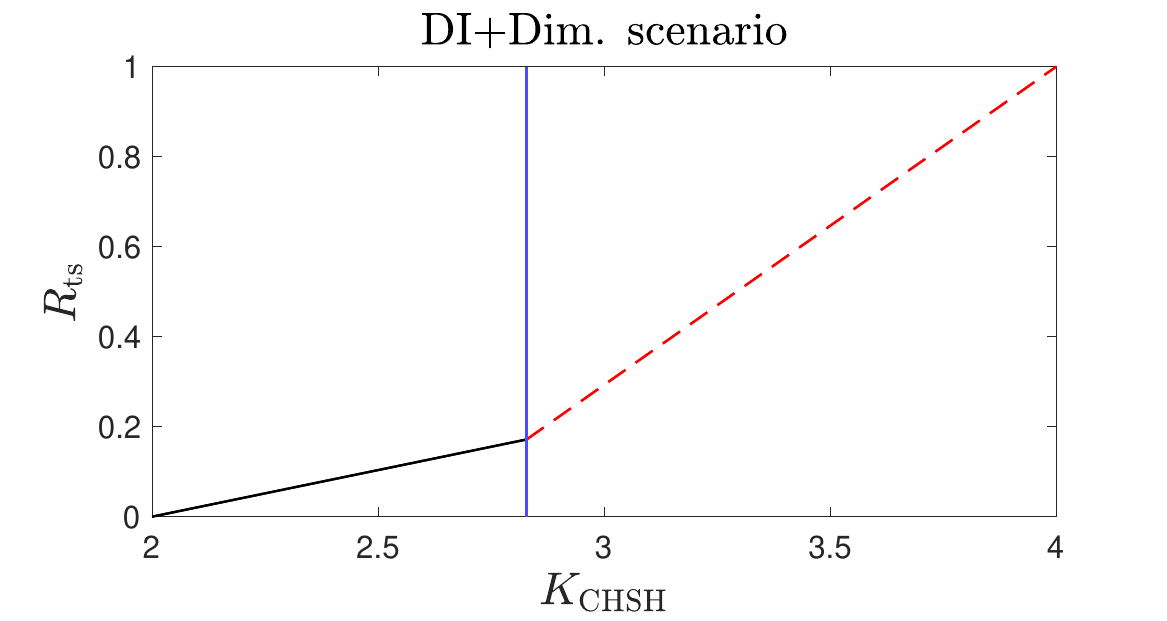}\\
\text{(b)}
\end{minipage}\\
\begin{minipage}[c]{.49\textwidth}
\includegraphics[width=9cm]{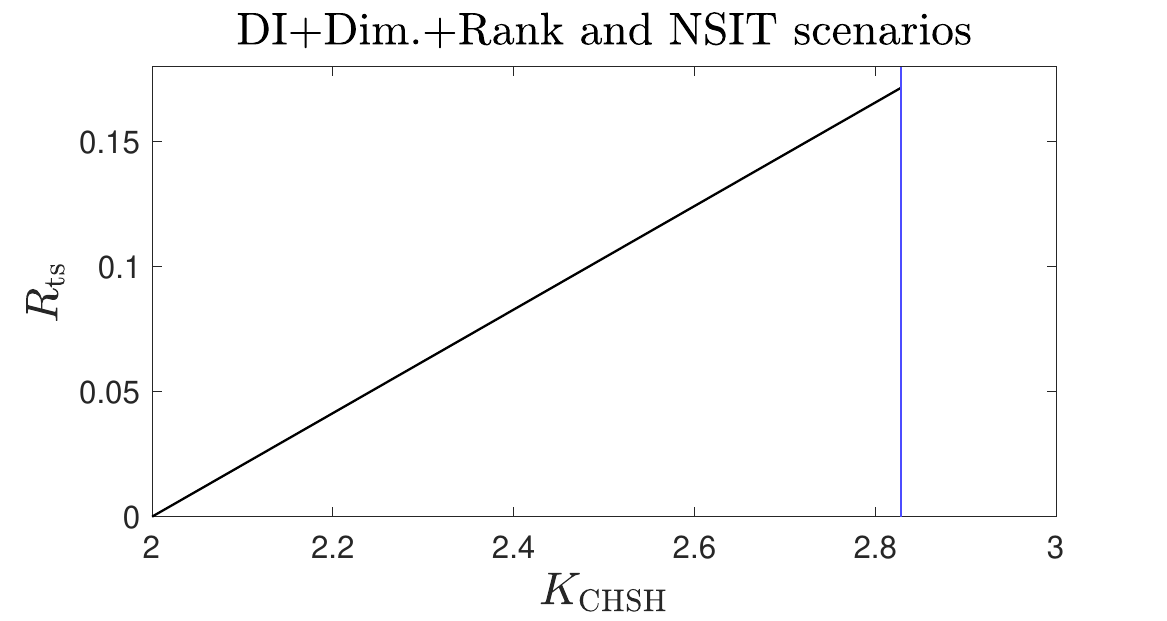}\\
\text{(c)}
\end{minipage}
\caption{
Given an observed quantum violation of the temporal CHSH value $K_{\rm CHSH}$, we estimate the minimal temporal steering robustness $R_{\rm ts}$ required to achieve the current value $K_{\rm CHSH}$.
(a) As can be seen in the figure, in the DI scenario, the bounds can be divided into two ranges of parameters: $2\leq K_{\rm CHSH}\leq 2\sqrt{2}$ and $2\sqrt{2}\leq K_{\rm CHSH}\leq 4$. The former is depicted in the black-solid curve, 
%remaining 
\je{signifying}
a noticeable gap (of the order of $10^{-3}$) with a straight line when the IMMs' level of the hierarchy achieves $5$. The latter is depicted in the red-dashed curve, converging to the straight line described by $R_{\rm ts}= (K_{\rm CHSH} - 2)(\sqrt{2}-1)/2$ when the IMMs' level of the hierarchy achieves $5$.
(b) The result of the DI+Dim. scenario is similar to the DI result. The difference is that the bounds converge to linearity, within the numerical precision, in both ranges of $2\leq K_{\rm CHSH}\leq 2\sqrt{2}$ ($1$st level of the IMMs) and $2\sqrt{2}\leq K_{\rm CHSH}\leq 4$ ($4$th level of the IMMs).
(c) The lower bounds on $R_{\rm ts}$ computed in the DI+Dim.+Rank and NSIT scenarios both match the straight line $R_{\rm ts}= (K_{\rm CHSH} - 2)(\sqrt{2}-1)/2$ within the numerical precision. The vertical blue lines in the three figures represent the value $2\sqrt{2}$, which is the maximal quantum violation in the DI+Dim.+Rank and NSIT scenarios.
}
\label{Fig_TSR_CHSH}
\end{figure*}

In this section, we show how to estimate the degree of temporal steerability (measured by the temporal steering robustness) in DI and semi-DI scenarios.
For the DI result, the method is similar to the work of Ref.~\cite{CBLC16}, where the authors have employed moment matrices induced by a bipartite system to quantify steerability. Here, we use the moment matrices 
induced by a single system to quantify temporal steerability. %First, 
\je{Consider} the \emph{temporal steering robustness}~\cite{HYKu16}, which is defined as the minimal ratio of the set of noisy post-measurement states $\JJaxs(\rho)$ one has to mix with $\IIaxs(\rho)$ before the mixture admits the hidden state model. That is, $R_{\rm ts} = \min \{t \Big\vert (\IIaxs(\rho)+t\JJaxs(\rho))/(1+t)=\sum_\lambda P(\lambda)P(a|x,\lambda)\sigma_\lambda \}$,
with $\JJaxs(\rho)\succeq 0$ and $\tr \sum_a\JJaxs(\rho)=1$.
\je{This gives}
\begin{equation}
\min_{\tilde{\sigma}_\lambda\succeq 0} ~\Big\{ \tr\sum_\lambda\tilde{\sigma}_\lambda - 1 \Big\vert \sum_\lambda \delta_{a,\lambda(x)}\tilde{\sigma}_\lambda - \IIaxs(\rho)\succeq 0\Big\},
\label{Eq_TSR_SDP}
\end{equation}
where each $\lambda$ is a vector whose $x$th element assigns a measurement outcome $a$, describing a deterministic strategy of observing outcome $a$ with choice $x$.  In a DI scenario, no assumption is 
made on $\IIaxs$ nor on $\rho$, therefore\je{,} the above semi-definite programming (SDP) cannot be computed. However, by applying the IMMs on the above SDP, some elements such as temporal correlations in the IMMs can be characterized, and for this reason the new SDP is solvable. The new constraints will be more relaxed (since we drop the characterization of $\IIaxs(\rho)$), 
therefore, the solution of the relaxed SDP will be a lower bound on $R_{\rm ts}$. The relaxed SDP is as follows
%We present the relaxed SDP and the numerical results in Appendix \ref{SecApp_DITSR}. For other semi-DI results, \je{we add the associated constraints}.
\begin{comment}
In the main text, the temporal steering robustness has been computed from the SDP 
as specified in
Eq.~\eqref{Eq_TSR_SDP} as
%which we write down here again:
\begin{equation}
\begin{aligned}
&R_{\rm ts} = \min_{\tilde{\sigma}_\lambda\succeq 0} ~\tr\sum_\lambda\tilde{\sigma}_\lambda - 1,\\
&\quad {\rm subject~to}\quad\sum_\lambda \delta_{a,\lambda(x)}\tilde{\sigma}_\lambda - \IIaxs(\rho)\succeq 0.
\end{aligned}
\end{equation}
A common tool to identify bounds is to make use of convex relaxations that give rise to
convex outer approximations of the feasible set in optimization problems. Such ideas of semi-definite
relaxations have first been
used in the context of quantum information science in Refs.~\cite{Doherty04,Hierarchies}.
As stated in the main text, by applying the IMMs on the above SDP, the new SDP features relaxed constraints, 
\end{comment}
%the form of which is as follows:
\begin{equation}
\begin{aligned}
\min \quad&\sum_\lambda\EBB[\sigma_\lambda]_{\openone,\openone} - 1
,\\
{\rm subject~to}\quad &\sum_\lambda\delta_{a,\lambda(x)}\EBB[\tilde{\sigma}_\lambda] - \EBB[\IIaxs(\rho)]\succeq 0,\\
&\EBB[\tilde{\sigma}_\lambda]\succeq 0,  \quad\EBB[\IIaxs(\rho)]\succeq 0,\\
&P(a,b|x,y) = P_{\rm obs}(a,b|x,y),
\end{aligned}
\end{equation}
where $\EBB[\sigma_\lambda]_{\openone,\openone}$ represents for $\tr(\sigma_\lambda S_j^\dag S_i)$ with $i,j$ being the indices for $S_i=S_j^\dag=\openone$. The solution of the above SDP is a DI lower bound on $R_{\rm ts}$.
For the DI+Dim.~result, the dimensional constraint is included in the above SDP of the form
\begin{equation}
\{\EBB[\IIaxs(\rho)]\}_{a,x}\in\mathcal{G}_d.
\end{equation}
For the DI+Dim.+Rank result, the additional rank constraint is included and the above equation is replaced by 
\begin{equation}
\{\EBB[\IIaxs(\rho)]\}_{a,x}\in\mathcal{G}_d^k.
\end{equation}
For the NSIT result, the above constraint is replaced by
\begin{eqnarray}
\sum_a \EBB[\IIaxs(\rho)] = \sum_a \EBB[\mathcal{I}_{a|x'}(\rho)]\quad\forall x\neq x'.
\end{eqnarray}
The results are shown in Fig.~\ref{Fig_TSR_CHSH}.

\section{Computing the successful probabilities in the $2\rightarrow 1$ and $3\rightarrow 1$ QRAC scenarios}\label{SecApp_ProbQRAC}

First, note that the post-measurement states depicted in our scenario (i.e., Fig.~\ref{Fig_scenario}) can be regarded as the set of states $\rho_{\vec{x}}$ prepared in QRAC scenario. As such, the formulation of moment matrices for $\rho_{\vec{x}}$ will be $\chi_{\vec{x}} = \sum_{i,j}\ket{i}\bra{j}\tr( \rho_{\vec{x}} S_j^\dag S_i )$. The accessible data $P(a',b'|x',y')$ in a general temporal scenario is associated with the average successful probability $P(b|\vec{x},y)$.  In fact, such a transformation can always be made by choosing $a'=x_0$, $x'=(x_1,x_2,\dots ,x_{n-1})$, $b'=b\in\{0,1\}$, and $y'=y\in\{0,1,\dots ,n-1\}$.  Consequently, for unknown states and measurements, the constraint of $\chi_{\vec{x}}\succeq 0$ naturally provides a characterization of quantum set of $P(b|\vec{x},y)$. For instance, the four prepared states $\rho_{x_0,x_1}$ in the $2\rightarrow 1$ scenario can be directly treated as the four post-measurement states $\{\mathcal{I}_{a'|x'}(\rho)\}_{a',x'}$ by choosing $a'=x_0$ and $x'=x_1$. The average successful probability for the $2\rightarrow 1$ scenario is given by $\mathcal{P}_{2\rightarrow 1} :=(1/8) \sum_{x_0,x_1,y}P(b=x_y|x_0,x_1,y)$ for $x_i,b,y\in\{0,1\}$. An upper bound on the maximum value of $\mathcal{P}_{2\rightarrow 1}$ for quantum strategies can be computed via 
\begin{equation}
\max\Big\{ \mathcal{P}_{2\rightarrow 1}
\Big\vert \chi_{x_0,x_1}\succeq 0,\quad
\{\chi_{x_0,x_1}\}_{x_0,x_1}\in\mathcal{G}_{d=2}^{k=1}\Big\}.
\label{Eq_QRAC2_SDP}
\end{equation}
%Note that 
We assume the measurements in the qubit-QRAC scenario to be projective, 
which is equal to requiring the POVMs be rank-one.
The result matches the quantum bound of 
\begin{equation}
\mathcal{P}_{2\rightarrow 1}^{\rm Q}:=\frac{1}{2}\left(1+\frac{1}{\sqrt{2}}\right)
\end{equation}
within the numerical precision for the first level of hierarchy of the IMMs (i.e., $\mathcal{S}=\{\openone, E_{1|1}, E_{1|2}\}$).
For the $3\rightarrow 1$ scenario, there are eight prepared states $\rho_{x_0,x_1,x_2}$ with $x_i\in\{0,1\}$. The correspondence with general temporal scenario can be made by choosing $a'=x_0$, $x'=(x_1,x_2)$, $b'=b\in\{0,1\}$, and $y'=y\in\{0,1,2\}$. The average successful probability is defined as 
\begin{equation}
\mathcal{P}_{3\rightarrow 1}:=\frac{1}{24} \sum_{x_0,x_1,x_2,y}P(b=x_y|x_0,x_1,x_2,y).
\end{equation}
Similarly with Eq.~\eqref{Eq_QRAC2_SDP}, an quantum upper bound on $\mathcal{P}_{3\rightarrow 1}$ can be computed. The result matches 
\begin{equation}
\mathcal{P}_{3\rightarrow 1}^{\rm Q}:=\frac{1}{2}\left(1+\frac{1}{\sqrt{3}}\right)  
\end{equation}
for the first level of hierarchy,  therefore, the bound is tight as well.

\section{Detailed derivation of the DI expression of the fidelity}\label{SecApp_DI_fidelity}
{\blu
In Ref.~\cite{Chen2021robustselftestingof}, the method of self-testing steerable assemblages \je{has been} proposed. Here, we apply their method on \je{notions of} self-testing preparation. The first step is to represent the completely positive map $\Lambda$ with the \emph{Choi-Jamio{\l}kowski} (CJ) matrix $\Omega$. Namely, 
\begin{equation}
\Lambda(\rho_{\vec{x}})=\tr_{\rm A}[\Omega(\rho_{\vec{x}}^\mathsf{T}\otimes\openone)] 
\end{equation}
or $[\Lambda(\rho_{\vec{x}})]^\mathsf{T}=\tr_{\rm A}[(\rho_{\vec{x}}\otimes\openone)\Omega^\mathsf{T}]$ with $\Omega:=(\mathsf{id}\otimes\Lambda)\ket{\phi^+}\bra{\phi^+}$, $\ket{\phi^+}:=\sum_i\ket{i}_{\rm A}\otimes\ket{i}_{{\rm A}'}$, and $\mathsf{id}$ being the identity map.  The second step is to choose an optimal quantum strategy which leads to the maximum of $P_{n\rightarrow 1}$.
In the main text, we consider the $2\rightarrow 1$ scenario, where the reference set of state vectors is composed of
\begin{eqnarray}
\ket{\psi_{0,0}^{\rm ref}}&=&c\ket{0}+s\ket{1},\\
 \ket{\psi_{1,0}^{\rm ref}}&=&s\ket{0}+c\ket{1}, \\
 \ket{\psi_{0,1}^{\rm ref}}&=&-c\ket{0}+s\ket{1}, \\
 \ket{\psi_{1,1}^{\rm ref}}&=&s\ket{0}-c\ket{1},
 \end{eqnarray}
  with $c:=\cos(\pi/8)$ and $s:=\sin(\pi/8)$.
This set is unitarily equivalent with the set $\{\ket{0},\ket{1},(\ket{0}+\ket{1})/\sqrt{2}, (\ket{0}-\ket{1})/\sqrt{2}\}$.
We choose the former since Bob's optimal observables are exactly the two Pauli observables: $B_0^{\rm ref}=\hat{Z}$ and $B_1^{\rm ref}=\hat{X}$.
Then, the map $\Lambda$ is chosen as the identity map. Consequently, the associated CJ matrix $\Omega$ is
\begin{equation}
\begin{aligned}
\Omega^\mathsf{T}&=\ket{0}\bra{0}\otimes\ket{0}\bra{0} + \ket{1}\bra{0}\otimes\ket{1}\bra{0} + \ket{0}\bra{1}\otimes\ket{0}\bra{1} +  \ket{1}\bra{1}\otimes\ket{1}\bra{1}\\
&=\frac{\openone+B_0^{\rm ref}}{2}\otimes\ket{0}\bra{0}+\frac{\openone-B_0^{\rm ref}}{2}B_1\frac{\openone+B_0^{\rm ref}}{2}\otimes\ket{1}\bra{0}\nonumber
+\frac{\openone+B_0^{\rm ref}}{2}B_1\frac{\openone-B_0^{\rm ref}}{2}\otimes\ket{0}\bra{1}+\frac{\openone-B_0^{\rm ref}}{2}\otimes\ket{1}\bra{1}.
\end{aligned}
\end{equation}
Apparently, there are many choices of the expression in the second line. Each choice can be regarded as an instance of completely positive and trace-preserving map $\Lambda$.
\je{Plugging the above expression} into $[\Lambda(\rho_{\vec{x}})]^\mathsf{T}=\tr_{\rm A}[(\rho_{\vec{x}}\otimes\openone)\Omega^\mathsf{T}]$, we have
\begin{equation}
\begin{aligned}
[\Lambda(\rho_{x_0, x_1})]^\mathsf{T}=
&\tr\left(\rho_{x_0, x_1}\frac{\openone+B_0^{\rm ref}}{2}\right)\ket{0}\bra{0}+
\tr\left(\rho_{x_0, x_1}\frac{\openone-B_0^{\rm ref}}{2}B_1^{\rm ref}\frac{\openone+B_0^{\rm ref}}{2}\right)\ket{1}\bra{0}\\
+&\tr\left(\rho_{x_0, x_1}\frac{\openone+B_0^{\rm ref}}{2}B_1^{\rm ref}\frac{\openone-B_0^{\rm ref}}{2}\right)\ket{0}\bra{1}+\tr\left(\rho_{x_0, x_1}\frac{\openone-B_0^{\rm ref}}{2}\right)\ket{1}\bra{1}.
\end{aligned}
\label{EqApp_Omega}
\end{equation}
Substituting the observables $B_0^{\rm ref}$ and $B_1^{\rm ref}$ with the POVMs elements $2E_{0|0}^{\rm ref}-\openone$ and $2E_{0|1}^{\rm ref}-\openone$, respectively, we have
\begin{equation}
\begin{aligned}
[\Lambda(\rho_{x_0, x_1})]^\mathsf{T}=&\tr\left(\rho_{x_0, x_1}E_{0|0}^{\rm ref} \right)\ket{0}\bra{0}+
\tr\left[\rho_{x_0, x_1}(2E_{0|1}^{\rm ref}E_{0|0}^{\rm ref}-2E_{0|0}^{\rm ref}E_{0|1}^{\rm ref}E_{0|0}^{\rm ref})\right]\ket{1}\bra{0}\\
+&\tr\left[\rho_{x_0, x_1}(2E_{0|0}^{\rm ref}E_{0|1}^{\rm ref}-2E_{0|0}^{\rm ref}E_{0|1}^{\rm ref}E_{0|0}^{\rm ref})\right]\ket{0}\bra{1}+\left[1-\tr(\rho_{x_0, x_1}E_{0|0}^{\rm ref})\right]\ket{1}\bra{1}.
\end{aligned}
\label{EqApp_Omega_2}
\end{equation}
Then we consider the definition of the fidelity between two sets of quantum states, namely~\cite{Liang19}, $F(\vec{\rho}, \vec{\sigma}):=\sum_{\vec{x}} F^{\rm UJ}(\rho_{\vec{x}}, \sigma_{\vec{x}})=\frac{1}{2^n}\sum_{\vec{x}}\tr(\rho_{\vec{x}}\sigma_{\vec{x}}),$ where $F^{\rm UJ}$ is the \emph{Uhlmann-Josza fidelity}~\cite{Uhlmann76,Jozsa1994} and the second equality holds when $\rho_{\vec{x}}$ or $\sigma_{\vec{x}}$ are pure.
With the above, we \je{find}%have
\begin{equation}
\begin{aligned}
&F(\{\rho_{x_0, x_1}^{\rm ref}\}, \{\Lambda(\rho_{x_0, x_1})\})\\
=&F(\{[\rho_{x_0, x_1}^{\rm ref}]^\mathsf{T}\}, \{[\Lambda(\rho_{x_0, x_1})]^\mathsf{T}\})\\
=&\frac{1}{4}\sum_{x_0,x_1}\tr\left(\rho_{x_0, x_1}^{\rm ref}\cdot[\Lambda(\rho_{x_0, x_1})]^\mathsf{T}\right)\\
=&\frac{1}{4}\Big[2+
(c^2-s^2)\tr(\rho_{0,0}E_{0|0}^{\rm ref})-4cs\tr(\rho_{0,0}E_{0|0}^{\rm ref}E_{0|1}^{\rm ref}E_{0|0}^{\rm ref})+2cs\tr(\rho_{0,0}E_{0|1}^{\rm ref}E_{0|0}^{\rm ref})+2cs\tr(\rho_{0,0}E_{0|0}^{\rm ref}E_{0|1}^{\rm ref})+\\
&\quad\quad\quad(s^2-c^2)\tr(\rho_{1,0}E_{0|0}^{\rm ref})-4cs\tr(\rho_{1,0}E_{0|0}^{\rm ref}E_{0|1}^{\rm ref}E_{0|0}^{\rm ref})+2cs\tr(\rho_{1,0}E_{0|1}^{\rm ref}E_{0|0}^{\rm ref})+2cs\tr(\rho_{1,0}E_{0|0}^{\rm ref}E_{0|1}^{\rm ref})+\\
&\quad\quad\quad(c^2-s^2)\tr(\rho_{0,1}E_{0|0}^{\rm ref})+4cs\tr(\rho_{0,1}E_{0|0}^{\rm ref}E_{0|1}^{\rm ref}E_{0|0}^{\rm ref})-2cs\tr(\rho_{0,1}E_{0|1}^{\rm ref}E_{0|0}^{\rm ref})-2cs\tr(\rho_{0,1}E_{0|0}^{\rm ref}E_{0|1}^{\rm ref})+\\
&\quad\quad\quad(s^2-c^2)\tr(\rho_{1,1}E_{0|0}^{\rm ref})+4cs\tr(\rho_{1,1}E_{0|0}^{\rm ref}E_{0|1}^{\rm ref}E_{0|0}^{\rm ref})-2cs\tr(\rho_{1,1}E_{0|1}^{\rm ref}E_{0|0}^{\rm ref})-2cs\tr(\rho_{1,1}E_{0|0}^{\rm ref}E_{0|1}^{\rm ref})\Big].
\end{aligned}
\end{equation}
Finally, relaxing the characterized states $\rho_{x_0, x_1}^{\rm ref}$ and POVMs $E_{b|y}^{\rm ref}$ to unknown ones $\rho_{x_0, x_1}$ and $E_{b|y}$, we have a DI expression of fidelity
\begin{equation}
\begin{aligned}
&F^{\rm DI}(\{\rho_{x_0, x_1}^{\rm ref}\}, \{\Lambda(\rho_{x_0, x_1})\})\\
=&\frac{1}{4}\Big[2+
(c^2-s^2)P(0|0,0,0)-4cs\tr(\rho_{0,0}E_{0|0}E_{0|1}E_{0|0})+2cs\tr(\rho_{0,0}E_{0|1}E_{0|0})+2cs\tr(\rho_{0,0}E_{0|0}E_{0|1})+\\
&\quad\quad\quad(s^2-c^2)P(0|1,0,0)-4cs\tr(\rho_{1,0}E_{0|0}E_{0|1}E_{0|0})+2cs\tr(\rho_{1,0}E_{0|1}E_{0|0})+2cs\tr(\rho_{1,0}E_{0|0}E_{0|1})+\\
&\quad\quad\quad(c^2-s^2)P(0|0,1,0)+4cs\tr(\rho_{0,1}E_{0|0}E_{0|1}E_{0|0})-2cs\tr(\rho_{0,1}E_{0|1}E_{0|0})-2cs\tr(\rho_{0,1}E_{0|0}E_{0|1})+\\
&\quad\quad\quad(s^2-c^2)P(0|1,1,0)+4cs\tr(\rho_{1,1}E_{0|0}E_{0|1}E_{0|0})-2cs\tr(\rho_{1,1}E_{0|1}E_{0|0})-2cs\tr(\rho_{1,1}E_{0|0}E_{0|1})\Big],
\end{aligned}
\end{equation}
where $P:=P(b|x_0,x_1,y)$.
}

\section{Classical fidelity}\label{SecApp_classical_fidelity}
In this section, we show how to compute the classical fidelity $f_c\approx 0.8536$ for the self-testing result plotted in Fig.~\ref{Fig_min_f_QRAC2to1}. The idea behind the definition of classical fidelity is straightforward: Given a reference set of state, we search for the best \emph{classical} set of states that gives the highest fidelity. That is,
\begin{equation}
f_c:=\max_{\vec{\rho}_{\rm c}} F(\vec{\rho}_{\rm ref}, \vec{\rho}_{\rm c})=\max_{\rho_{\vec{x}}^{\rm c}}\frac{1}{4}\sum_{x_0,x_1}\tr(\rho_{x_0, x_1}^{\rm ref}\rho_{x_0, x_1}^{\rm c}),
\label{EqApp_Fc}
\end{equation}
where $\vec{\rho}_{\rm c}$ denotes a set of classical set of states. In Ref.~\cite{Tavakoli18b}, the authors have fairly defined a classical set of states: \je{This is the} %The 
set of states 
whose elements are all diagonal states, i.e., 
\begin{equation}
\rho_{\vec{x}}^{\rm c}:=\sum_i \alpha_{\vec{x},i}\ket{i}\bra{i}~~\forall\vec{x},\quad {\rm with}~~\sum_i \alpha_{\vec{x},i}=1~~\forall\vec{x}~~{\rm and}~~\alpha_{\vec{x},i}\geq 0~~\forall\vec{x},i,
\end{equation}
where $\{\ket{i}\}$ is an orthonormal basis and $\alpha_{\vec{x},i}$ are some real numbers. With this, Eq.~\eqref{EqApp_Fc} can be computed via the linear program
\begin{equation}
\max_{\{\alpha_{x_0,x_1,i}\}}\Big\{ \frac{1}{4}\sum_{x_0,x_1,i}\alpha_{x_0,x_1,i}\bra{i}\rho_{x_0, x_1}^{\rm ref}\ket{i} \Big\vert  \sum_i \alpha_{x_0,x_1,i}=1~~\forall x_0,x_1~~{\rm and}~~\alpha_{x_0,x_1,i}\geq 0~~\forall x_0,x_1,i \Big\},
\end{equation}
and one can obtain $f_c\approx 0.8536$.

\section{Generalization to a multiple-time scenario and to many-body systems}\label{SecApp_generalization}
\subsection{Generalization to a multiple-time scenario}
\newtext{
Consider we apply multiple measurements on a single system at $N$ many points of time. The quantum description of each measurement (except for the last one) labelled by $x_k$ at time $t_k$ is described by an instrument $\{\mathcal{J}^k_{a_k|x_k}\}_{a_k}$, with 
the collection of $a_k$ being the measurement outcomes. The measurements at the final time are described by POVMs $E_{a_N|x_N}$, i.e., they satisfy $E_{a_N|x_N}\succeq 0$ for all $a_N,x_N$ and $\sum_{a_N} E_{a_N|x_N} = \mathbb{1}$ for all $x_N$. The quantum temporal correlations can be obtained by the Born rule:
\begin{equation}
P(a_1,a_2,\dots, a_N|x_1,x_2,\dots, x_N) = \tr\Big\{
E_{a_N|x_N} \Big[
\mathcal{J}^{N-1}_{a_{N-1}|x_{N-1}}\Big(
\mathcal{J}^{N-2}_{a_{N-2}|x_{N-2}}\Big(
...\Big(
\mathcal{J}^{1}_{a_1|x_1}(\rho)
\Big)...
\Big)
\Big)
\Big]
\Big\},
\end{equation}
where $\rho$ is the initial state. Given that the composition of instruments is a valid instrument, we can rewrite the above equation as
\begin{equation}
P(a_1,a_2,\dots, a_N|x_1,x_2,\dots, x_N) = \tr\Big[
E_{a_N|x_N}\cdot \mathcal{I}_{a_{N-1},a_{N-2},\dots, 1|x_{N-1},x_{N-2},\dots, 1}(\rho)
\Big]\quad\forall a_1,\dots, a_{N-1},x_1,\dots, x_{N-1},
\end{equation}
where
\begin{equation}
\mathcal{I}_{a_{N-1},a_{N-2},\dots, 1|x_{N-1},x_{N-2},\dots, 1}:=
\mathcal{J}^{N-1}_{a_{N-1}|x_{N-1}}\circ
\mathcal{J}^{N-2}_{a_{N-1}|x_{N-2}}\circ
...\circ
\mathcal{J}^{1}_{a_1|x_1}\quad\forall a_1,\dots, a_{N-1},x_1,\dots, x_{N-1}.
\end{equation}

As before, let the initial state $\rho$ act on the set $\mathcal{L}(\mathcal{H}_{\text{A}_1})$, then we have $\mathcal{J}^{1}_{a_1|x_1}$: $\mathcal{L}(\mathcal{H}_{\text{A}_1})\rightarrow \mathcal{L}(\mathcal{H}_{\text{A}_2})$,\dots, $\mathcal{J}^{k}_{a_k|x_k}$: $\mathcal{L}(\mathcal{H}_{\text{A}_k})\rightarrow \mathcal{L}(\mathcal{H}_{\text{A}_{k+1}})$,\dots, $\mathcal{J}^{N-1}_{a_{N-1}|x_{N-1}}$: $\mathcal{L}(\mathcal{H}_{\text{A}_{N-1}})\rightarrow \mathcal{L}(\mathcal{H}_{\text{B}_1})$ and $E_{a_N|x_N}\in \mathcal{L}(\mathcal{H}_{\text{B}_1})$. This yields $\mathcal{I}_{a_{N-1},a_{N-2},\dots, 1|x_{N-1},x_{N-2},\dots, 1}: \mathcal{L}(\mathcal{H}_{\text{A}_1})\rightarrow \mathcal{L}(\mathcal{H}_{\text{B}_1})$. With these, we can introduce a completely positive map $\EBB: \mathcal{L}(\mathcal{H}_{\text{B}_1})\rightarrow \mathcal{L}(\mathcal{H}_{\bar{\text{B}}_1})$ and define the $\ell$th level instrument moment matrices as (c.f., Eq.~(2) in the main text)
\begin{equation}
\chi_{a|x}^{(\ell)} := \EBB[\mathcal{I}_{a_{N-1},a_{N-2},\dots, 1|x_{N-1},x_{N-2},\dots, 1}(\rho)] = \sum_{i,j}\ket{i}\bra{j}\tr\left[ \mathcal{I}_{a_{N-1},a_{N-2},\dots, 1|x_{N-1},x_{N-2},\dots, 1}(\rho) S_j^\dag S_i \right],
\end{equation}
where $\{S_i\}$ is defined as $\openone\cup\mathcal{S}^{(1)}\cup\mathcal{S}^{(2)}\cup \dots \cup\mathcal{S}^{(\ell)} $, with $\mathcal{S}^{(\ell)}$ being composed of the $\ell$th-order products of the operators in the set $\{E_{a_N|x_N}\}$.

The instrument moment matrices $\chi_{a|x}^{(\ell)} $ are positive semi-definite by construction if all the components in the setting, namely, $\rho$, $\mathcal{I}_{a_{N-1},a_{N-2},\dots, 1|x_{N-1},x_{N-2},\dots, 1}$, and $E_{a_N|x_N}$ have valid quantum descriptions. Therefore, the positive semi-definite constraints $\chi_{a|x}^{(\ell)}\succeq 0 $ characterize that the observed correlation $P(a_1,a_2,\dots, a_N|x_1,x_2,\dots, x_N)$ admits a quantum realization.
}

\subsection{Generalization to many-body systems}
\newtext{
Let us consider that the initial state $\rho$ is prepared in a multipartite or many-body quantum state composed of $N$ many subsystems. Namely,
\begin{equation}
\rho\in\mathcal{L}(\mathcal{H}_{\text{A}_1}\otimes\mathcal{H}_{\text{A}_2}...\otimes \mathcal{H}_{\text{A}_N}).
\end{equation}
Then, the state is sent into an instrument $\mathcal{I}_{a_1,a_2,\dots, a_N|x_1,x_2,\dots, x_N} :=\mathcal{I}_{\vec{a}|\vec{x}}: \mathcal{L}(\mathcal{H}_{\text{A}_1}\otimes\mathcal{H}_{\text{A}_2}...\otimes \mathcal{H}_{\text{A}_N})\rightarrow \mathcal{L}(\mathcal{H}_{\text{B}_1}\otimes\mathcal{H}_{\text{B}_2}...\otimes \mathcal{H}_{\text{B}_N})$, where $a_k$ are the measurement results and $x_k$ are the measurement choices of the $k$th subsystem. The post-measurement state $\mathcal{I}_{\vec{a}|\vec{x}}(\rho)$ is measured by measurements described by POVMs $E_{b_1,b_2,\dots, b_N|y_1,y_2,\dots, y_N}:=E_{\vec{b}|\vec{y}}\in \mathcal{L}(\mathcal{H}_{\text{B}_1}\otimes\mathcal{H}_{\text{B}_2}...\otimes \mathcal{H}_{\text{B}_N})$, where $b_k$ are measurement outcomes and $y_k$ are measurement choices on the $k$th subsystem. Note that $\mathcal{I}_{\vec{a}|\vec{x}}$ and $E_{\vec{b}|\vec{y}}$ are now written in a general case but they can be in some special forms when introducing explicit assumptions. For instance, if the measurements are all separable, then we have
\begin{equation}
\mathcal{I}_{\vec{a}|\vec{x}} = \mathcal{I}_{a_1|x_1}\otimes\mathcal{I}_{a_2|x_2}\otimes...\otimes\mathcal{I}_{a_N|x_N}
\end{equation}
and
\begin{equation}
E_{\vec{b}|\vec{y}} = E_{b_1|y_1}\otimes E_{b_2|b_2}\otimes...\otimes E_{b_N|y_N}.
\end{equation}
The quantum temporal correlations are obtained by the Born rule according to
\begin{equation}
P(\vec{a},\vec{b}|\vec{x},\vec{y}):=P(a_1,\dots, a_N,b_1,\dots, b_N|x_1,\dots, x_N,y_1,\dots, y_N) = \tr\big[ E_{\vec{b}|\vec{y}} \mathcal{I}_{\vec{a}|\vec{x}}(\rho) \big].
\end{equation}
To define the instrument moment matrices, we now introduce completely positive maps $\EBB: \mathcal{L}(\mathcal{H}_{\text{B}_1}\otimes\mathcal{H}_{\text{B}_2}...\otimes \mathcal{H}_{\text{B}_N})\rightarrow \mathcal{L}(\mathcal{H}_{\bar{\text{B}}_1}\otimes\mathcal{H}_{\bar{\text{B}}_2}...\otimes \mathcal{H}_{\bar{\text{B}}_N})$ and define the $\ell$th level IMMs as
\begin{equation}
\chi_{a|x}^{(\ell)} := \EBB[\mathcal{I}_{\vec{a}|\vec{x}}(\rho)] = \sum_{i,j}\ket{i}\bra{j}\tr\left[ \mathcal{I}_{\vec{a}|\vec{x}}(\rho) S_j^\dag S_i \right],
\end{equation}
where $\{S_i\}$ is defined as $\openone\cup\mathcal{S}^{(1)}\cup\mathcal{S}^{(2)}\cup \dots \cup\mathcal{S}^{(\ell)} $ and $\mathcal{S}^{(\ell)}$ is composed of the $\ell$th-order products of the operators in the set $\{E_{\vec{b}|\vec{y}}\}$.

By construction, the IMMs $\chi_{a|x}^{(\ell)}$ are positive semi-definite if the initial state $\rho$ and the instrument $\mathcal{I}_{\vec{a}|\vec{x}}$ admit quantum descriptions. Therefore, the positive semi-definite constraints $\chi_{a|x}^{(\ell)}\succeq$ are a characterization of the set of quantum temporal correlations $P(\vec{a},\vec{b}|\vec{x},\vec{y})$ for many-body systems.}

%\clearpage
%%%%%%%%%%%%%%%%%%%%%%%%%%%%%%%%%%%%%%%%
%\bibliographystyle{abbrv}
%\bibliographystyle{unsrt}
%\bibliographystyle{alpha}

%%%%%%%%%%%%%%%%%%%%%%%%%%%%%%%%%%%%%%%%

\end{document}